\documentstyle[twocolumn]{mn}

\title[Turbulent galactic dynamos]
{Can the turbulent galactic dynamo generate large-scale magnetic fields ? }

\author[K. Subramanian]
       { Kandaswamy Subramanian\thanks{On leave from 
       National Centre for Radio Astrophysics, TIFR,
        Poona University Campus, Ganeshkind, Pune 411007. India.
        email: ksub@astr.maps.susx.ac.uk, kandu@gmrt.ernet.in } \\
         Astronomy Centre, University of Sussex, Falmer, 
         Brighton BN1 9QH , UK.\\ Princeton University Observatory, Peyton
         Hall, Princeton, NJ 08544. USA. \\ 
         NCRA, TIFR,
        Poona University Campus, Ganeshkind, Pune 411007. India}

\date{}

\pagerange{\pageref{firstpage}--\pageref{lastpage}}
\pubyear{1997}

\begin{document}

\maketitle

\label{firstpage}

\begin{abstract}

Large-scale magnetic fields in galaxies are thought to be generated by
a turbulent dynamo. However the same turbulence also
leads to a small-scale dynamo which generates
magnetic noise at a more rapid rate. 
The efficiency of the large-scale dynamo depends on how
this noise saturates. We examine this issue taking into 
account ambipolar drift, which obtains in a galaxy with
significant neutral gas. We argue that, (1) the small-scale 
dynamo generated field does not fill the volume, but is
concentrated into intermittent rope like structures. 
The flux ropes are curved on the turbulent eddy scales. Their
thickness is set by the diffusive scale determined by the
effective ambipolar diffusion; (2) For a largely neutral galactic
gas, the small-scale dynamo saturates,
due to inefficient random stretching, 
when the peak field in a flux rope 
has grown to a few times the equipartition value; 
(3) The average energy density in the saturated 
small-scale field is sub equipartition, 
since it does not fill the volume;
(4) Such fields neither drain significant 
energy from the turbulence nor convert eddy motion of the turbulence 
on the outer scale into wavelike motion. The 
diffusive effects needed for the large-scale dynamo 
operation are then preserved until the large-scale field
itself grows to near equipartition levels.

\end{abstract}


\section{Introduction}
 
The origin of ordered, large-scale galactic magnetic fields remains 
a challenging problem.
Magnetic fields in galaxies have strengths of order $10^{-6} G$,
and are coherent on scales of several kpc (Beck {\it et al.} 1996).  
These fields can arise, in principle, 
by dynamo generation, from a weak, but nonzero, seed field 
$\sim 10^{-19} - 10^{-23} G$,
if the galactic dynamo can operate efficiently enough to exponentiate the
field by a factor $\sim 30 - 40$ (cf. Moffat 1978, Parker 1979,
Zeldovich {\it et al.} 1983).
However the capacity of presently known turbulent
dynamo mechanisms to produce the observed galactic fields
has been debated (Cattaneo and Vainshtein 1991, 
Vainshtein and Rosner 1991, Kulsrud and Anderson 1992, 
Brandenberg 1994).

It has been argued 
that magnetic noise in the form of small-scale fields, builds up 
much faster than the mean field in a turbulent flow. 
Magnetic noise can result from the tangling of the large-scale
field by the turbulence, or the operation of a small-scale
dynamo. The dominant source, when one starts from weak large-scale
seed fields, is the operation of a small-scale dynamo. 
Turbulence with a large enough magnetic Reynolds number (MRN),
even when mirror-symmetric on average,
generically leads to an exponential growth of 
fields correlated on the turbulent eddy scales, 
{\it independent} of the large-scale field. This growth occurs on the
the turbulent eddy turnover timescale which 
is much smaller than the
time scale for the growth of the mean field. Therefore
the kinematic dynamo paradigm will become invalid
long before the large-scale field has grown
anywhere near the observed levels.
Kulsrud and Anderson (KA) (1992) reach the 
somewhat bold conclusion that the 
galactic field must be of primordial origin! Although there are 
a number of mechanisms to produce small seed magnetic fields
(Rees 1987, 1994, Ratra 1992, Subramanian 1995,
Subramanian, Narasimha and Chitre, 1995 and references therein ),
there is as yet no compelling mechanism to produce a field, anywhere near 
that required by the primeval hypothesis.

This problem may disappear if the small-scale dynamo
generated fields can saturate, 
due to non-linear back reaction effects, in a manner which does not
destroy the ability of turbulent motions to amplify the large-scale
field. We examine whether this can indeed happen.
Ideally one has to consider the MHD dynamo, 
where both the induction equation for the magnetic field, and the Euler 
equation for the velocity field, are solved simultaneously in 
a self consistent fashion. However this is a formidable task at present,
even for numerical simulation. In this paper we take a first look at 
this difficult non-linear problem, in a simpler fashion by
isolating the ingredients needed for
the small-scale dynamo action and investigating
under what conditions these get suppressed. 

Galaxies have a significant neutral gaseous component. 
As magnetic fields grow, the
Lorentz force on the charged component will cause a 
slippage between it and the neutrals. Its magnitude
is determined by the friction between the components due 
to ion-neutral collisions. This drift, called ambipolar drift 
(Mestel and Spitzer 1956, Spitzer 1978, Draine 1980, 1986, Zweibel 1986),
is one important non-linear feedback on both the small
and large-scale dynamos, by
the generated magnetic field.
In another paper (Subramanian 1997 ;
Paper II, in preparation), we give a derivation 
of the equations for both mean field and 
the magnetic correlations, incorporating the effects of ambipolar
drift. We also give there the solution of these equations in several contexts. 
Some pertinent results of this work will be quoted here, where needed.
We will see below, that 
the presence of neutrals, may be very important,
in leading to a saturated state for 
the small-scale dynamo, which
preserves large-scale dynamo action.

In section 2, we begin by introducing the turbulent
galactic dynamo. In section 3, we summarise the
properties of the kinematic small-scale dynamo
action in Kolmogorov type
turbulence and point out the problem they raise.
The influence of ambipolar drift
on the small-scale dynamo is considered in section 4.
Section 5 considers the the back reaction effects 
and saturation of the small-scale dynamo, 
due to the Lorentz forces acting on the fluid as a whole.
The last section contains a discussion and summary of our results.
We argue that, in galaxies, magnetic noise
generated by the small-scale dynamo, 
may indeed saturate in a fashion which preserves large 
scale dynamo action.

\section{ The turbulent galactic dynamo }

Spiral galaxies are differentially rotating systems.
The magnetic flux is to a 
large extent frozen into the fluid and so any radial
component of the magnetic field will be efficiently wound 
up and amplified to produce a toroidal component of the field.
But this only results in a linear amplification of the field: 
to obtain the observed 
galactic fields starting from small seed fields  
we must find a way to generate the radial 
components of the field in the galaxy from the toroidal one.
If this can be done, the field can grow exponentially and 
one has a dynamo.

The standard picture involves 
the effects of cyclonic turbulence in the 
galactic gas. The galactic interstellar medium is assumed 
to be turbulent, due for example to the effects of supernovae randomly 
exploding in different regions. In a rotating, stratified 
(in density and pressure) medium,
like a disk galaxy, such turbulence becomes cyclonic and
acquires a net helicity. 
Isotropic and homogeneous turbulence with helicity, in the 
presence of a large-scale magnetic field, ${\bf B}_0$, leads to an 
extra electromotive force of the form 
$ {\bf E} = \alpha {\bf B}_0 - \eta_t {\bf \nabla } \times {\bf B}_0$, 
where $\alpha $ depends on the helical part of the turbulence 
and $\eta_t $ is the turbulent diffusion which depends on the 
non helical part of the turbulent velocity correlation function
(Krause $\&$ Radler 1980, Moffat 1978, Parker 1979).
It should be noted that both the alpha effect, and turbulent diffusion,
depend crucially on the diffusive (random walk) property of 
fluid motion (cf. Field 1996). So, 
if for some reason (see below) the fluid motion becomes
wavelike, then the alpha effect and turbulent diffusion will
be suppressed (Chandran 1996).

The induction equation, with the extra turbulent component of 
the electric field, a prescribed large-scale velocity field, 
can have exponentially growing solutions for the large-scale field.
These have been 
studied extensively in the literature 
(cf.Ruzmaikin, Shukurov $\&$ Sokoloff 1988, Mestel $\&$ Subramanian 1991,
 Beck et al. 1996 for a recent review). 
It had been assumed in most earlier works that the turbulent 
velocities do not get affected by Lorentz forces - 
until the mean large-scale 
field builds up sufficiently. However, this does not turn out to
be valid due to the more rapid build up of magnetic noise 
compared to the value of the mean field, a problem to which we now turn.

\section{The Kinematic small-scale dynamo and the problem of magnetic noise}

Split up the magnetic field, ${\bf B} = {\bf B}_0 
+ \delta{\bf B}$, into a mean field $ {\bf B}_0$ and a 
fluctuating component  $\delta{\bf B}$. Here, the mean field,
${\bf B}_0 = < {\bf B} >$, is defined
either as a spatial average over scales larger than the turbulent
eddy scales or, more correctly, as an ensemble average.
Kazantsev (1968) was the first to show that, even purely 
mirror-symmetric turbulence, leads to dynamo amplification of
the small-scale fluctuating fields, 
for a sufficiently large magnetic Reynolds number
. Some of the subsequent work is summarised
by Zeldovich {\it et al.} (1983). The statistical properties
of the small-scale field are most clearly expressed in terms of the 
magnetic correlation function 
$M_{ij}(r,t) = <\delta B^i({\bf x},t) \delta B^j({\bf y},t) > $,
where $r = \vert {\bf x} - {\bf y} \vert$. 
Small scale dynamo action is described by the
evolution of the longitudinal component
$M_L(r,t) = r^ir^j M_{ij}/r^2$, 
where $r^i = x^i - y^i$.

In the case when the
turbulent velocity, say ${\bf v}_T$, has a delta function correlation
in time (Markovian), it is relatively straightforward
to derive the evolution equation for $M_{ij}$ (cf. Kazantsev 1968,
Vainshtein $\&$ Kichatinov 1986, Paper II).
Suppose we also assume ${\bf v}_T$ 
to be an isotropic, homogeneous,
Gaussian random velocity field with 
zero mean. Specify its two point correlation function by
$ <v^i_T({\bf x},t)v^j_T({\bf y},s)>  = T^{ij}(r) \delta (t - s)$, with
\begin{equation}
T^{ij}(r) =
T_{NN}[\delta^{ij} -({r^i r^j \over r^2})] +
 T_{LL}({r^i r^j \over r^2}) +
C\epsilon_{ijf} r^f .
\end{equation}
Here, $T_{LL}(r)$ and $T_{NN}(r)$ are the 
longitudinal and transverse correlation functions for the velocity 
field and $C(r)$ represents 
the helical part of the velocity correlations.
(cf. Landau and Lifshitz 1987). If ${\bf v}_T$ is assumed to be
divergence-free, then, 
\begin{equation}
T_{NN}(r) =  {1 \over 2r} {\partial \over \partial r} ( r^2 T_{LL}(r) ) .
\end{equation}

The evolution of $M_L$ is given by
\begin{equation}
{\partial M_L \over \partial t} = {2\over r^4}{\partial \over \partial r}
(r^4 \kappa {\partial M_L \over \partial r}) + G M_L ,
\label{mleq}
\end{equation}
where we have defined 
\begin{eqnarray}
&&\kappa = \eta + T_{LL}(0) - T_{LL}(r) , \nonumber\\ 
&&G = - 4\left[ {d\over dr}({T_{NN}\over r}) + {1\over r^2} 
{d\over dr}(rT_{LL})
\right] .
\end{eqnarray}
The term involving $\kappa$ in equation \ (\ref{mleq})
represents the effects of diffusion. 
The diffusion coefficient includes the effects of 
microscopic diffusion ($\eta$) and a scale-dependent 
turbulent diffusion, ($T_{LL}(0)-T_{LL}(r)$).
The term proportional to $G(r)$, allows  
for the rapid generation of magnetic fluctuations, through 
shearing action and the existence of
a small-scale dynamo
{\it independent of the large-scale field}.
[ In the evolution equation for $M_L$ given above, we have neglected its
coupling to the helical part of the magnetic correlation
due to a scale-dependent alpha effect. In the
galactic context this has a negligible effect.
We have also neglected the subdominant coupling to the
large-scale field (see Paper II).]

Suppose $V$ and $L$ are the velocity
and correlation lengths of the dominant energy carrying
eddies of the turbulence, which is assumed to have a Kolmogorov 
energy spectrum. For numerical estimates we generally
take $V = 10 km s^{-1}$ and $L = 100 pc$. 
For Kolmogorov turbulence, the eddy
velocity at any scale $l$, is $v_l \propto l^{1/3}$,
in the inertial range. The turbulence is cut off at
a scale, say $l_c \approx L R_e^{-3/4}$, where $R_e = VL/ \nu$ is the
fluid Reynolds number and $\nu$ is the kinematic viscosity.
We will be considering a largely
neutral galactic gas and for this $\nu$ is dominated by 
the neutral contribution.
We take the neutral-neutral collision
to be dominated by H-H collisons with a cross section 
$\sigma_{H-H} \sim 10^{-16} cm^{-2}$, leading to a
kinematic viscosity $\nu \sim v_{th} (1 / n_H \sigma_{H-H})$. 
For a thermal velocity $v_{th} \sim 10 km s^{-1}$ and a
neutral hydrogen number density $n_H \sim 1 cm^{-3}$, we have
$\nu \sim 10^{22} cm^2 s^{-1}$, so 
\begin{equation}
R_e = {VL \over \nu } \approx 3 \times 10^{4} V_{10} L_{100}
\label{rey}
\end{equation}
where $V_{10} = (V/ 10 {\rm km s}^{-1})$ and $L_{100} =
(L/ 100 {\rm pc})$.
The magnetic Reynolds 
number at the outer scale of the turbulence is defined to be
$R_m = (VL/\eta)$. For the Spitzer value of the resistivity
$\eta = 10^7 (T/10^4K)^{-3/2} cm^2 s^{-1}$, with turbulence parameters
as above, $R_m = 3 \times 10^{19}$. 
Since $v_l \propto l^{1/3}$, the magnetic Reynolds number associated
with eddies of size $l$ scales as $R_m(l) = v_l l/\eta = R_m (l/L)^{4/3}$.

The evolution of $M_L(r,t)$ has been studied in detail by
several authors (Zeldovich {\it et al} 1983 and references therein)
for the case of when $T_{LL}(r)$ has a single scale. We also study
in Paper II, using WKBJ technique, 
the evolution of $M_L(r,t)$ for a model Kolmogorov
type turbulence. We look at the properties
of exponentially growing modes, $M_L(r,t) = M_n(r) e^{\Gamma_n t}$, 
which are regular at the origin, and tend to zero as $r \to \infty$.
Some results of this work and the earlier works of Kazantsev (1968) and
Zeldovich {\it et al.} (1983) pertinent to the present context,
are summarised here.

\begin{itemize}

\item There is a critical value for the magnetic Reynolds number (MRN):
$R_m = R_c \approx 60$, for
the excitation of the small-scale dynamo. Above this critical MRN
the small-scale dynamo can lead to an exponential growth of the
fluctuating field correlated on a scale $L$.
Further, the equations determining
$R_c$ are the same if we replace $(L,R_m)$ by $(l,R_m(l))$. Therefore,
the critical MRN for excitation of a
mode concentrated around $r\sim l$ is also $R_m(l) = R_c$, as
expected from the scale invariance in the inertial range.

\end{itemize}
 
In the galactic context $R_m >> R_c$; in fact, one also has
$R_m(l_c) = v_c l_c/\eta = R_m/R_e >> 1$. (Here $v_c$ is the
eddy velocity at the cut-off scale).
Hence, small-scale dynamo action
excites modes correlated on all scales from the cut-off
scale $l_c$ to the external scale $L$ of the turbulence.

\begin{itemize}

\item Due to small-scale dynamo action, the fluctuating field, 
tangled on a scale $l$, grows exponentially on the corresponding
eddy turnover time scale, 
with a growth rate $\Gamma_l \sim v_l/l$. 
Since $v_l \sim l^{1/3}$, the growth rate is $\Gamma_l \propto l^{-2/3}$,
and so increases with decreasing $l$.
In the galactic context, with
$R_m(l_c) = R_m/R_e >> R_c$, 
the small-scale fields tangled at the cutoff scale
grow more rapidly than any of the large-scale modes.

\item The WKBJ analysis gives a growth rate
$\Gamma_c = (v_c/l_c) [5/4 - c_0 ({\rm ln}(R_m/R_e))^{-2} ]$
with $c_0 = \pi^2/12$ for the fastest mode. Also the growth
rate for the small-scale dynamo is only weakly 
(logarithmically) dependent on $R_m$, provided $R_m$ is large enough.

\item For the parameters adopted above, we have  
$\Gamma_L^{-1} \sim L/V \sim 10^7 {\rm yr}$.
For the modes tangled at the cut-off scale, this time is
$\Gamma_c^{-1} \sim 10^4 {\rm yr}$. These times
are much smaller 
than the time scale for the growth of the large-scale field.

\end{itemize}

The spatial structure of the dynamo-generated small-scale
field is important in determining how the small-scale
dynamo saturates. To examine the spatial structure for various 
eigenmodes of the small-scale dynamo, 
it is more instructive to consider the function $w(r,t) = 
<\delta {\bf B}({\bf x},t) .\delta {\bf B}({\bf y},t)> $, 
which measures the correlated 
dot product of the fluctating field ($w(0) = 
<\delta {\bf B}^2 >$ ).   
Firstly there is a general constraint that can be placed
on $w(r)$. Since the fluctuating field is 
divergence free, we have (Kleeorin {\it et al.} 1986, Paper II),
\begin{equation}
w(r,t) = {1\over r^2} {d\over dr}\left [ r^3 M_L \right ] ,
\label{wr}
\end{equation}
so 
\begin{equation}
\int_0^{\infty} w(r) r^2 dr = \int_0^{\infty} 
{d\over dr}\left [ r^3 M_L \right ]  = 0 , 
\label{consw}
\end{equation}
 since $M_L$ is regular at the origin and vanishes faster than
$r^{-3}$ as $r \to \infty$. Therefore
the curve $r^2w(r)$ should have zero area
under it. Since $w(0) = < (\delta{\bf B})^2>$, 
$w$ is positive near the origin. And
the fluctuating field points in the same direction for small
separation. As one goes to larger values of $r$, there must then 
values of $r$, say $r \sim d$, where $w(r)$ becomes negative.
For such values of $r$, the field at the origin and at a separation
$d$ are pointing in opposite directions on the average. 
This can be intepreted 
as indicating that the field lines, 
on the average are curved on the scale $d$.

\begin{figure*}
\centering
\begin{picture}(300,250)
\includegraphics{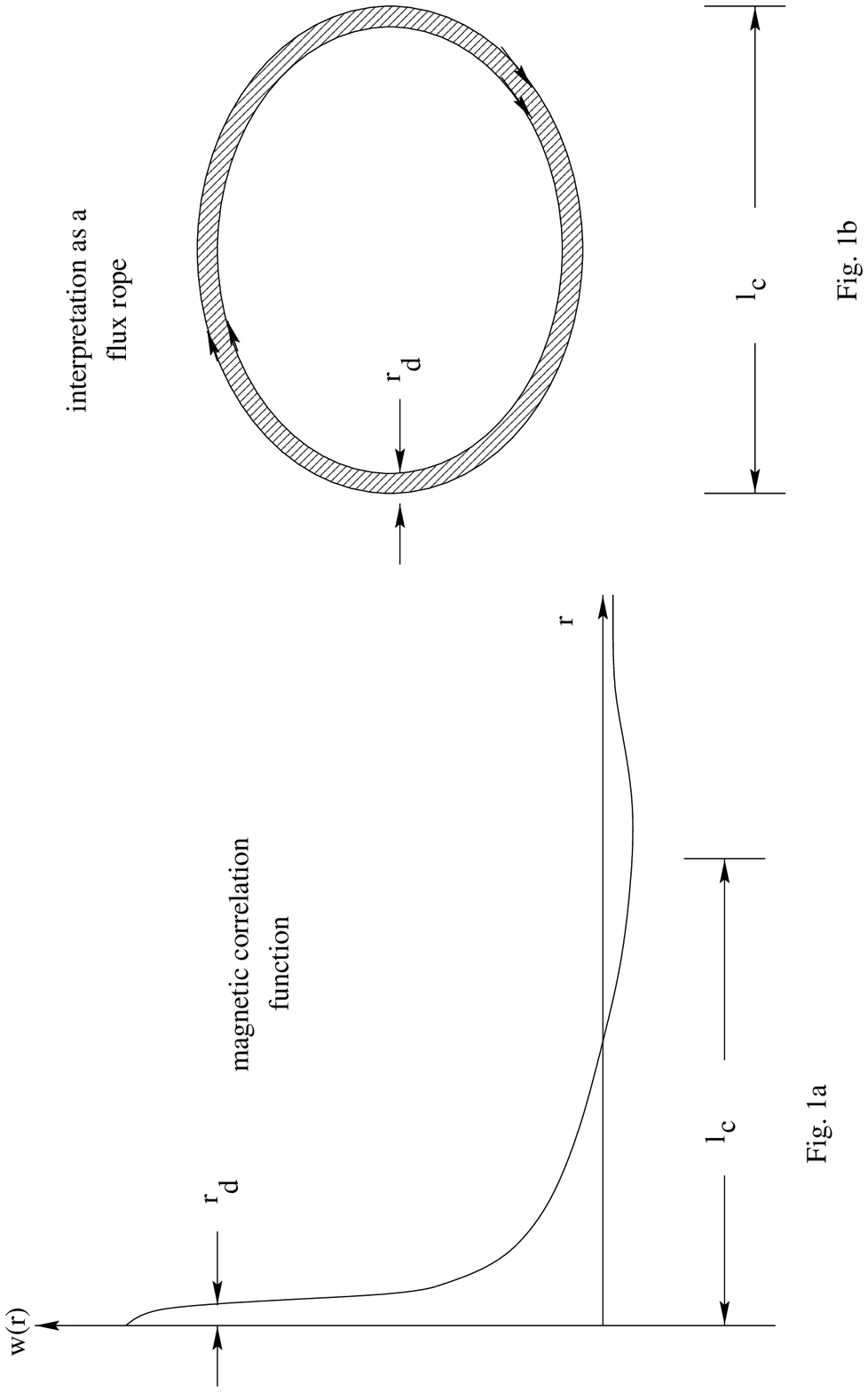}
\end{picture}
%\begin{figure}
%\vspace*{300pt}
\caption{(a) Schematic representation of the
the magnetic correlation function $w(r)$ for the fastest
growing mode in the kinematic regime.
The fact that $w(r)$ is positive
at small $r$ and concentrated
into a region of scale $r_d$ near the origin, implies
that the field points in the same direction on
the average for separations smaller than the diffusive scale
$r_d$. The anti-correlation tail, with $w(r) < 0$, is related
to the vanishing of field divergence; that 
for points separated by the eddy scale $l_c$,
the field points in opposite directions, on average.
(b) The pictorial intepretation of $w(r)$ 
as a flux rope, with $r_d$ as the rope thickness and
$l_c$ as curvature scale of the flux rope.}
\end{figure*}

\begin{itemize}

\item In the case $R_m/R_e >> 1$, $w(r)$ is strongly peaked 
within a region $r =r_d \approx l_c (R_m/R_e)^{-1/2}$ about the origin, 
for all the modes. Note that $r_d$ is the diffusive scale 
satisfying the condition $\eta/r_d^2 \sim v_c/l_c$. 

\item For the most rapidly growing mode, $w(r)$ 
changes sign accross $r \sim l_c$ and rapidly decays
with increasing $r/l_c$. 

\item For slower growing modes, with 
$\Gamma_l \sim v_l/l$, $w(r)$ extends up to $r \sim l$
after which it decays exponentially.

\end{itemize}

We should point out that a detailed 
analysis of the eigenfunctions, 
for the simple case when the longitudinal
velocity correlation function has only a single scale,
can be found in Kleeorin et. al. (1986). Their analysis
is also applicable to the mode near the cut-off scale in
Kolmogorov type turbulence. Further, these authors elaborate
on a pictorial intepretation of the correlation function, in terms of the
Zeldovich rope-dynamo (cf. Zeldovich {\it et al.} 1983). 
We have shown, schematically, $w(r)$ 
for the the fastest growing mode
in Figure 1a. and its pictorial intepretation interms of
a flux rope in Figure 1b. For the fastest growing mode there is only 
one node where $w(r)=0$.
For higher order modes with smaller growth rates, 
several nodes for $w(r)$ will occur
and can be intepreted interms of several
scales on which the field is curved (cf. Kleeorin {\it et al.} 1986,
see also Ruzmaikin {\it et al.} 1989 ). The extent of $w(r)$,
(after which it decays exponentially), gives the largest 
scale on which the flux rope is curved.

\begin{itemize}

\item If one adopts this intepretation of
the $w(r)$ for the various modes, 
the small-scale field can be 
thought of as being concentrated in rope like structures with 
thickness of order the diffusive scale $r_d << l$ 
and curved on a scale up to $\sim l$ for  
modes extending up to $r\sim l$.

\end{itemize}

To end this section we give a qualitative picture
of the mechanism for the dynamo growth of small-scale fields.
When the field starts from an arbitrary initial
configuration, in the kinematic stage, 
the initial field growth, is just due to random
stretching by the turbulence, together with flux freezing. 
Eddies of scale $l$ stretch and tangle the field on corresponding 
scales. The field grows at this stage because
the stretching of a "flux rope" leads to a decrease in its 
cross section (due to near incompressibility of the flow), and hence 
from flux freezing an increase in the field strength. As the field
grows, the magnetic field gets concentrated into smaller and
smaller scales until diffusion comes into play.
 
Consider to begin with the effect 
of eddies at the cut-off scale, $l_c$, of the turbulence. 
The initial amplification due to purely stretching of
the field by these eddies stops when
the field has been concentrated into a small enough
scale $r_d$, such that the rate of diffusion across 
$r_d$ becomes comparable to the 
stretching rate by these eddies.
This is when 
\begin{equation}
\eta/r_d^2 \sim v_c/l_c .
\label{thickl}
\end{equation}
This gives $r_d \sim l_c/R_m^{1/2}(l_c)$.
We will refer to $r_d$ as the 
thickness of the flux rope curved on a scale $l_c$.
Further growth of the field can 
only be achieved by the operation of the small-scale
dynamo, which exponentiates the field for a MRN above $R_c$.
This is explicitly demonstrated by the solutions for
the kinematic small-scale dynamo discussed above 
and can be thought of as the operation of the Zeldovich -
Stretch-Twist-Fold - rope dynamo at random locations.

For galactic gas with 
$R_m/R_e >> 1$, even the eddies at the cut-off scale 
have an MRN greater than the critical value
needed for dynamo action. These eddies exponentiate the
field at a growth rate $v_c/l_c$. 
Larger eddies of scale $l > l_c$ also lead to stretching,
twisting and folding of the field at a slower rate, $v_l/l$. This leads to
dynamo growth of fields tangled at scale $l$, with
a slower growth rate $v_l/l$. The 
field curved on a scale $l$ is also, at the kinematic stage,
chopped up further by smaller scale eddies (a scale dependent turbulent
diffusion ) until its energy can be dissipated by 
microscopic diffusion at scale $r_d$. 
So, on any flux rope of length $l$, one has smaller scale
wiggles until the diffusive scale $r_d$.  

We emphasise that the time scale for 
mean-field growth is $\sim 10^9 yrs$, 
of order a few rotation time scales of the 
disk, and is much larger than the time scale for the growth of
the fluctuating field ( $\Gamma_L^{-1} \sim 10^7 {\rm yr}$).
Hence, the operation of the small-scale dynamo will imply
that the magnetic field 
is rapidly dominated by the fluctuating component.
Chandran (1996) has shown that 
the presence of small-scale magnetic fields could
change hydrodynamic turbulence into magneto-elastic waves, with
a phase velocity $\bar v_A$, where $\bar v_A^2$ is
two thirds of the magnetic energy per unit mass.
So, if the energy density in the
small-scale magnetic noise builds up to
equipartition levels, the fluid motions could become
predominantly wavelike, with a wave period of
order the eddy turn-over time. This could then
lead to a reduced alpha effect and turbulent 
diffusion. Unless the build up of magnetic
noise is curbed in a way which leaves the turbulence
still having a diffusive property, large-scale dynamo action
will be severely affected. We now turn to the effect of ambipolar drift
and the possible ways in which the small-scale dynamo saturates.

\section{ The effect of ambipolar drift}

In a partially ionised medium the magnetic field evolution is
governed by the induction equation
\begin{equation}
(\partial {\bf B}/ \partial t) =
{\bf \nabla } \times ( {\bf v}_i \times {\bf B} - 
 \eta {\bf \nabla } \times {\bf B}), 
\label{basici}
\end{equation} 
where ${\bf v}_i$ the velocity of the ionic component of the fluid.
The ions experience the Lorentz force due 
to the magnetic field. This will cause them to drift with respect to
the neutral component of the fluid. If the ion-neutral 
collisions are rapid enough, one can assume that the Lorentz force on 
the ions is balanced by their friction with the neutrals. Under this 
approximation, the Euler equation for the ions reduces to :
\begin{equation}
\rho_i \nu_{in} ({\bf v}_i -{\bf v}_n )  
= {({\bf \nabla } \times {\bf B}) \times {\bf B} \over 4\pi },
\end{equation}
where $\rho_i$ is the mass density of ions, $\nu_{in}$ the ion-neutral 
collision frequency and ${\bf v}_n $ the velocity of the neutral particles.

The ion-neutral elastic scattering frequency is given by
$\nu_{in} = \rho_n <\sigma v>/(m_i + m_n)$, where $\rho_n$ is the
neutral fluid density, and $m_i, m_n$ are the ion and neutral particle masses
(cf. Mestel and Spitzer 1956, Draine 1980, 1986). 
We will assume that the galaxy had
very nearly primordial composition in its early stage of evolution:
then the ions are mostly just protons and the neutrals are
mostly hydrogen atoms. Elastic scattering 
occurs with the ion polarising the neutral atom, and interacting
with the induced dipole. 
For $H-H^+$ interactions, in addition to elastic scattering, 
there can also be charge exchange reactions, 
which increase the ion-neutral cross-section.
Draine (1980) adopts the maximum of these two rates, and gives
a momentum transfer rate coefficient of $<\sigma v> \approx 
3.2 \times 10^{-9} cm^{3} s^{-1}$ for $v < 2 km s^{-1}$,
and $<\sigma v> \approx 2.0 \times 10^{-9} (v/ km s^{-1})^{0.73} cm^{3} s^{-1}$ 
for $ 2 km s^{-1} < v < 1000 km s^{-1}$. In the galactic disk, 
we expect the gas to have a temperature $T < 10^4 K$ with 
$ v \sim 10 km s^{-1}$ and so $<\sigma v> (H-H^+) \sim 
10.74 \times 10^{-9} cm^{3} s^{-1}$. 
Further, the interaction with helium atoms will not give
a significant addition to the collision rate because
the polarisability of helium is less due to its symmetry and
helium is 4 times heavier than hydrogen. However, helium will 
contribute about $25 \%$ of the total density of the fluid.
Let $\rho_{H^+}$, $\rho_H$ be the proton and hydrogen densities and
$n_i = \rho_{H^+} /m_H$. We then have 
\begin{equation}
\rho_i \nu_{in} = {\rho_{H^+} \rho_{H} <\sigma v>_{H-H^+} \over 2 m_H} 
= n_{i} \rho_n < \sigma v>_{eff} 
\label{incoll}
\end{equation}
with $<\sigma v>_{eff} \sim 4 \times 10^{-9} cm^{3} s^{-1}$.

For the evolution of the small-scale field, we show in Paper II,
that ambipolar drift adds to the diffusion
coefficient, $\kappa$, a term proportional 
to the energy density in the fluctuating fields.
This changes $\eta$ to an effective value 
\begin{equation}
\eta_{ambi} = \eta + { w(0,t) \over 6\pi\rho_i \nu_{in}} = 
\eta + {< \delta {\bf B} . \delta {\bf B}> \over 6\pi\rho_i \nu_{in}} ,
\end{equation} 
and replaces $\kappa$ in equation (\ref{mleq}) for $M_L(r,t)$,  
by a new $\kappa_N = \eta_{ambi} + T_{LL}(0) - 
T_{LL}(r)$.
One can define an effective magnetic Reynolds number, incorporating
the effect of ambipolar drift, for
fluid motion on any scale of the turbulence by
\begin{equation}
R_{ambi}(l) = {v_l l \over \eta_{ambi}} 
= { v_l l  6\pi\rho_i \nu_{in} \over < \delta {\bf B} . \delta {\bf B}> } ,
\label{rmeffi}
\end{equation}
where $v_l = (l/L)^{1/3} V$ as before.

As the energy density in the fluctuating field 
increases, $ R_{ambi}(l)$ decreases.
Firstly, this makes it easier for the field energy to reach the
diffusive scales, from a general 
initial configuration. After this stage,
the initial amplification due to purely stretching stops, and
further growth of the field can 
only be achieved by the operation of the small-scale
dynamo.
If, as the field grows, 
$R_{ambi}(L)$ decreases sufficiently, 
a stationary state with $\partial M_L /\partial t = 0$ could, 
in principle, be achieved.
In such a state, $M_L$ is independent of time.
So, the condition on the critical MRN for the stationary
state to be reached, will
be identical to that obtained in the kinematic stage.
That is, if $R_{ambi}(L)$ decreases to a value $R_c \sim 100$, 
dynamo action will stop completely. 

However, for galactic turbulence,  
\begin{equation}
R_{ambi}(l)  = {1 \over f(l) } {3\rho_i \nu_{in} l \over 2\rho_n v_l} =
{Q(l) \over f(l)} ,
\label{rambi}
\end{equation}
where $f(l) = B_l^2/(4\pi \rho_n v_l^2)$ is the ratio of the local magnetic
energy density of a flux rope curved on scale $l$, 
to the turbulent energy density $ \rho_n v_l^2/2$
associated with eddies of scale $l$.
Using the value of $\nu_{in}$ as determined
in Eq.\ (\ref{incoll}) and putting in numerical
values we get 
\begin{equation}
Q(l) = {3\rho_i \nu_{in} l \over 2\rho_n v_l} 
\sim 1.8 \times 10^{4} n_{-2}
({l \over L})^{2/3} L_{100} V_{10}^{-1}  
\label{rambii}
\end{equation}
where $n_{-2} = (n_i / 10^{-2} cm^{-3}) $, and 
we have assumed a 
Kolmogorov scaling for the turbulent velocity
fluctuations.

One can see from Eq.\ (\ref{rambi}) - (\ref{rambii}) 
that, for typical parameters
associated with galactic turbulence, the MRN incorporating
ambipolar drift is likely to remain much larger than $R_c$ for
most scales of the turbulence, even when the field energy density
becomes comparable to the equipartition value. So 
ambipolar drift by itself cannot saturate the small
scale dynamo. Rather,
one expects the field to continue to grow rapidly,
even taking into account ambipolar drift.
Note also that the growth rates for the small-scale 
dynamo generally depends only weakly
on the MRN, provided the MRN is much larger than $R_c$.
(see section 3, Kleeorin et al. 1986, Ruzmaikin {\it et al.} 1989).
Therefore, we still expect the small-scale dynamo-generated field to grow 
almost exponentially on the eddy turn around time scale,
as long as $R_{ambi} >> R_c$.

The spatial structure of the fluctuating field will also
remain ropy, as argued in section 3, as long as $R_{ambi} >> R_c$.
However, as the field strength in a flux rope, curved on a
scale $> l$, grows to near equipartition with the turbulent energy
associated with eddies smaller than $l$, these 
smaller scale eddies would no longer be effective in
causing "turbulent diffusion" of the larger scale field.
So the thickness $r_d(l)$
of a flux rope curved on a scale $l$ 
is determined eventually by demanding that 
the {\it ambipolar diffusion } 
timescale across $r_d(l)$ becomes comparable to the 
stretching timescale $l/v_l$; that is,   
\begin{equation}
r_d^2(l)/\eta_{ambi} \sim l/v_l .
\label{thick}
\end{equation}
This determines the thickness to be $r_d(l) \sim l/R_{ambi}^{1/2}(l)$.
Since $R_{ambi}(l) >> 1$, we expect 
flux ropes to remain relatively thin
with a thickness $\sim l/R_{ambi}^{1/2}(l) << l$, even taking account of 
the ambipolar drift. 

In summary, we have argued here that, in galaxies, 
the small-scale dynamo continues to 
exponentiate the field fluctuations
even in the presence of ambipolar drift. This fluctuating field 
however does not fill the volume but is
concentrated into intermittent rope like structures. 
The ropes are curved on the turbulent eddy scale and their
thickness is set by the diffusive scale $r_d(l)$ determined by the
effective ambipolar diffusion. 

We have to consider how other non-linear feedback processes
could limit the growth of this small-scale dynamo generated field. 

\section{ Saturation of the small-scale dynamo }

\subsection{Inefficient random stretching and damping }

The first of these restraining processes 
is the reduction in the efficiency of stretching
of a flux rope as the field in the rope, say $B_p$, grows in strength.
A turbulent eddy of scale $l$ produces a correlated winding up of the field 
for a time of order $l/v_l$. Note that this is just a consequence
of the induction $\nabla \times ({\bf v} \times {\bf B})$ term and
velocity shear. No dynamics is involved. However, 
suppose the growing tension component 
of the Lorentz force on the
flux rope tangled on this scale, can untangle (or straighten) the
rope and damp away its wrinkle, 
on a comparable timescale. The random stretching of
the flux rope by these eddies will be suppressed and the small
scale dynamo will not operate efficiently. 

To estimate the straightening time-scale, say $t_s$, we have to look
at the dynamics of the flux rope.
The motion of the flux rope through the surrounding medium 
is influenced not only by the tension force but also by friction.
(We are neglecting, for now, the effects of 
rotation and gravity). Suppose $v_s$ is the velocity associated
with the untangling or straightening of the field in  
the flux rope. If one were to ignore the effect of friction,
we would have $v_s \sim v_A = (B_p^2/ 4 \pi \rho_n)^{1/2}$. However, the
effect of drag leads to a smaller "terminal" velocity, which
can be estimated by equating the 
tension force to the drag on the flux rope. Also, the work
done by the rope against the frictional
drag leads to a damping of the energy associated
with the wrinkle of the flux rope  on the
straightening timescale, $l/v_s$, which is comparable
to the rapid eddy turnover time when $v_s \sim v_l$.
We illustrate schematically, this process of flux rope
straightening and damping in Figure 2a.

\begin{figure*}
\centering
\begin{picture}(300,250)
\includegraphics{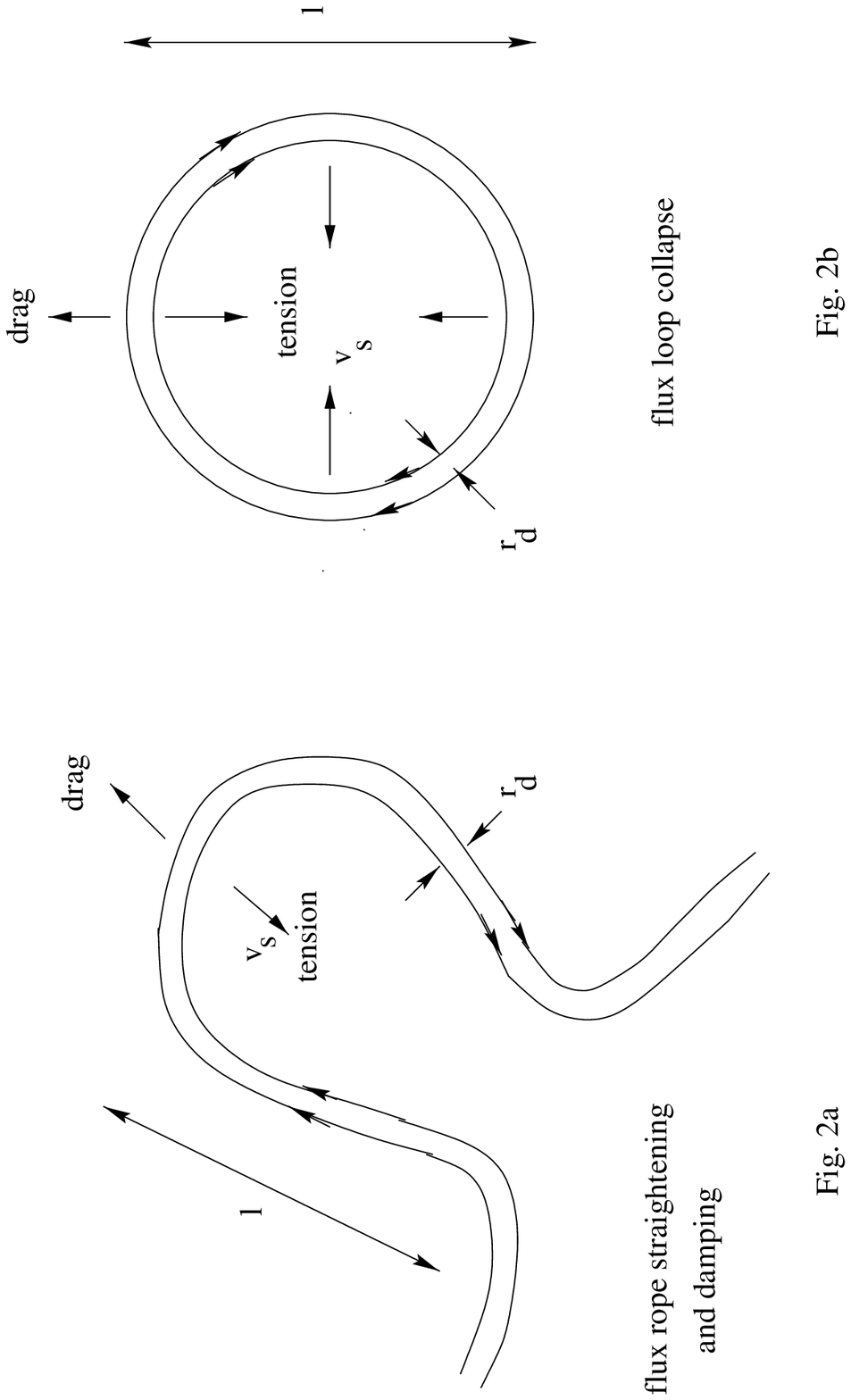}
\end{picture}
%\begin{figure}
%\vspace{300pt}
\caption{(a) A schematic illustration of a curved flux
rope, with a radius of curvature $l$ and thickness $r_d$.
Magnetic tension acts to straighten the rope and 
aerodynamic drag damps the magnetic energy associated with 
the wrinkle in the rope. This leads to inefficient random
stretching when $v_s$ becomes comparable to $v_l$.
(b) The collapse of flux loops is illustrated
schematically. Such collapse results in an irreversible
removal of small-scale magnetic noise and limits the
value of $N$.}
\end{figure*}

The drag force per unit length 
on the flux rope is  $\sim C_d \rho_n v_s^2  r_d/2 $ 
(cf. Parker 1979 Eq. 8.59). The coefficient $C_d$ depends on the 
fluid Reynolds number on the scale of the radius of the flux rope,
that is on $R_e^{rad} = v_s r_d/\nu$. Note that the drag formalism
is not well developed for the case when the external medium is itself
turbulent (cf. Parker 1979 Section 8.7). One expects
eddies with a scale smaller than the rope 
radius to have a different effect compared to larger eddies.
In our case, $r_d << l$, and so one expects
momentum transfer by eddies with scale $< r_d$ to be
subdominant compared to the turbulent drag induced by larger 
eddies. Here we adopt the drag formula given above,
with the assumption that the complications mentioned above
do not drastically change the results.

Equating the magnetic tension component of the Lorentz
force in the rope and drag then gives
\begin{equation}
{B_p^2\over 4\pi (l/2) } (\pi r_d^2)  \sim  
{C_d \over 2} \rho_n v_s^2 r_d .
\label{ambst}
\end{equation} 
Here, we have taken the curvature radius associated with the
field tangled on scale $l$ to be $l/2$. 
The dynamo action on these scales 
will be affected when $t_s \sim l/v_s \sim l/v_l$ or when 
$v_s \sim v_l$. Since,
in this case, the flux rope is able to straighten
and damp (due to friction) its wrinkle (curvature) 
on a timescale comparable to
the stretching timescale $l/v_l$. 
So, the random stretching of the field by eddies of scale $l$,
will become inefficient when the peak field has grown to
a value found by substituting $v_s \sim v_l$ in Eq. (\ref{ambst}).
Straight forward rearrangment
of the various quantities in Eq. \ ( \ref{ambst} )
then implies an upper limit on the magnetic field in the flux rope, of
\begin{equation}
f_p(l)= { B_p^2/8\pi \over (\rho_n v_l^2/2) } \approx  
{C_d  \over 4\pi} {l \over r_d}.
\label{pkbcon}
\end{equation}
Since we expect $ r_d < l$, the peak field in the rope can exceed
equipartition values.

When ambipolar drift is causing the flux rope to thicken, we
have $l/r_d = R_{ambi}^{1/2}(l)$, which itself depends on the magnetic 
field strength in the rope. Further, the drag co-efficient, $C_d$, 
depends on $R_e^{rad}$, so on $r_d$, and therefore
implicitly on the field in the rope.

The value of $C_d$ decreases from about 30 at
$R_e^{rad} \sim 1$ to 4 at $R_e^{rad} \sim 10$,  1 at $R_e^{rad} \sim 10^2$
and ranges between 1-0.1 for larger values (cf. Parker 1979 Sec.8).
For our problem, an adequate estimate of the 
drag co-efficient is obtained by taking $C_d \sim 10/\sqrt{R_e^{rad}(r_d)}$.
Taking $v_s \sim v_l$, we have $R_e^{rad} = R_e(l) (r_d/l)$.
Using $l/r_d = R_{ambi}^{1/2}(l)$, and re-arranging, we
obtain
\begin{eqnarray}
f_p(l) &\approx& ({10\over 4\pi})^{4/7}
{Q^{3/7}(l) \over R_e^{2/7}(l)} \nonumber\\ &\approx& 3.1 \ 
n_{-2}^{3/7} ({l\over L})^{-2/21} L_{100}^{1/7}
V_{10}^{-5/7}
\label{seclim}
\end{eqnarray}
One can now go back and estimate the
Reynolds number $R_e^{rad}$. We get 
\begin{equation}
R_e^{rad} \sim 392 (l/L)^{19/21} n_{-2}^{-2/7}V_{10}^{3/2}
L_{100}^{1/2}.
\end{equation}
At the cut off scale $l_c$, $R_e^{rad} \sim 0.4$,
while at the outer scale $L$, $R_e^{rad} \sim 392$. So 
the approximation used for the drag co-efficient should be reasonably 
accurate.

We see from Eq. (\ref{seclim}) , 
that when the field in the flux rope of scale $l$ grows to a few times
the equipartition field associated with eddies on that scale, 
the random stretching of the rope will become inefficient,
limiting its further growth by dynamo action.
Since larger eddies carry larger energy, this also implies that the
the saturated value of the field  in a flux rope, 
tangled on a scale $l$ will be larger for larger $l$. 

The average energy density in the field
tangled on the scale $l$, however need not exceed the energy
density of the turbulent eddies because of the ropiness of the field.
Since the ropes have a thickness
of order $r_d$ and are curved an a scale $l$, the average
energy density contributed by a flux rope to a sphere of radius $l$ is
\begin{equation}
E_B(l) \sim {B_p^2 \over 8\pi} { \pi r_d^2 \pi l \over (4\pi/3)l^3}
=  {B_p^2 \over 8\pi} {3 \pi \over 4} {r_d^2 \over l^2} .
\end{equation}
Note that, although $B_p^2 \propto (l/r_d)$, and can exceed
equipartition value, the average energy density of a flux rope
in a correlation
volume $E_B(l) \propto B_p^2 (r_d/l)^2 \propto (r_d/l)$.
This can be much smaller than equipartition because $r_d/l << 1$.

In principle, the length of the flux rope tangled on a 
scale $l$ may be larger than the value $\pi l$ assumed above, 
and equal to say $N \pi l$. In this case,
the ratio of the average magnetic
energy density of the field tangled on scale $l$ to the 
turbulent energy on the same scale is 
\begin{equation}
F(l) = {E_B(l)\over (\rho_nv_l^2/2)}
 \sim N ({3C_d \over 16}) ({ r_d \over l}) .
\end{equation}
Putting in numerical values, we have 
\begin{equation}
F(l) =  0.9 \times 10^{-3} N
n_{-2}^{-1/4} ({l\over L})^{-5/6}L_{100}^{-3/4}
V_{10}^{-1/4}
\label{energy}  
\end{equation}

We can also ask if the energy dissipated in ambipolar drift and
frictional drag is comparable
to the turbulent power in the case when the small
scale field saturates due to the stretching constraint.
Since the ambipolar drift rate across the rope is
comparable to the stretching rate, $v_l/l$, the 
power dissipated by ambipolar drift is $(\eta_{ambi}/r_d^2(l)) E_B(l) \sim 
E_B(l)(v_l/l)$. The power lost due to frictional
drag is also $\sim E_B(l) (l/v_l)$. So the ratio of the total power 
dissipated to the turbulent power is
\begin{equation}
{P_D \over P_T} = 2{E_B(l) (v_l/l) \over (\rho_n v_l^2/2)(v_l/l)} \sim 2F(l)
\label{powst}
\end{equation}

Both the energy density in the flux ropes, and the energy
dissipated due to ambipolar drift and friction
depend on how large $N$ can get,
or how many flux ropes of scale $l$ are packed into a correlation
volume of radius $l$, in the final saturated state.

\subsection{The limiting effect due to flux loop collapse}
 
An important process, which limits
$N$ from becoming too
large, enters as the peak field in the flux rope increases
to the  value given by Eq. (\ref{pkbcon}).
As one tries to pack more and more flux ropes curved
on a scale $l$ into a volume of same scale, the 
probability that the rope intersects itself, or another
co-habiting rope, increases. This will result in loops
of flux of radius $l$. Also, note that the the very process
by which the small-scale dynamo may
operate, viz. via the stretching-twisting-folding actions
associated with the Zeldovich rope dynamo, will generically
result in flux loops of scale $l$.

These loops of flux, when not being stretched, can 
collapse to a small radius (cf. Deluca, Fisher $\&$ Patten 1993,
Vishniac 1995).This is schematically illustrated in Figure 2b. 
The time-scale for the collapse is of order
$l/v_s $, which will be $\sim l/v_l$, the eddy turnover time,
since $v_s \sim v_l$ when the peak field in the rope
is given by Eq. (\ref{pkbcon}). 
 The collapse of the 
loop will be halted when it has shrunk to a radius comparable
to its thickness.
At this stage, the effects of diffusion
could convert most of the remaining energy 
in the loop into kinetic energy and 
heat. This process results in the irreversible
removal of energy from the small-scale magnetic field
tangled on scale $l$ at a rate $v_l/l$, comparable to its rate
of build up by turbulent stretching, after the field has grown
sufficiently. 
A dynamical equilibrium for N can then result, 
whereby  "old" loops of flux, which have peak flux
given by Eq.  (\ref{pkbcon}), collapse and are destroyed
at a rate $v_l/l$, to be replaced by newly created loops, at the same rate, 
which are just reaching the saturated value of the peak flux.
This picture of dynamical equilibrium leads one to
conjecture that, in the saturated state, the average length
of flux ropes tangled on scale
$l$, in a volume of radius $l$, 
cannot grow much larger than $\pi l$. 
That is the value of $N$, can
not grow much larger than unity.

Given that $N$ is not too large, one can see from Eq. (\ref{powst}) 
that the power dissipated in ambipolar drift is much smaller 
than the turbulent power. This occurs for
all scales except near the cut-off scale. 
In fact, one only needs $N$ less than
about $100$ for the eddies at the energy-carrying scale to be left
unaffected (undamped) by the growth of the small-scale field.
Also, from Eq. (\ref{energy}), 
one can see that the average energy density of the 
generated small-scale field is much smaller than the average
energy density in the turbulence.  
So, any wave-like
motion induced by the presence of the small-scale 
field (cf. Chandran 1996) will have 
a period larger than the eddy turn around time. This implies
that such tangled small-scale fields do not change the
diffusive nature of the turbulence. For these reasons, 
the large-scale dynamo can still operate to generate the mean 
field. 

\subsection{ The limit on the field in flux ropes 
due to the external pressure of the gas}

There is one caveat to the above discussion. We have assumed that
the field in the rope can grow sufficiently so that the
tension in the rope begins to play an important role in the
rope dynamics. However, there is an upper limit to 
the growth of the magnetic field in the ropes, from the effect of
its magnetic pressure on the dynamo process.
Due to the increasing importance of this pressure,
stretching of field lines can lead to a partial decrease in
fluid density in the ropes rather than a decrease
in the rope-cross section and the associated increase
in the rope magnetic field (cf. Vishniac 1995). 
An upper limit to the
magnetic pressure in the ropes is given by the
external pressure $P_{ext}$. This implies that the field in the 
rope, is limited to 
\begin{equation}
B_p < (8\pi P_{ext})^{1/2}.  
\label{finlim}
\end{equation}
Whether the field in the ropes will be limited by inefficient 
random stretching as given by Eq. (\ref{seclim})
or by external pressure (Eq. (\ref{finlim}))
will depend on the parameters of the
problem. The rope field will be given by the
lower of the limiting fields implied by Eqs. (\ref{seclim})
and (\ref{finlim}).

The total pressure of the interstellar medium in a galaxy 
could contain a number of components; a thermal component,
pressure due to turbulence itself and possibly due to non-thermal
"cosmic- rays".
The ratio of the gas pressure to the 
turbulent energy density is 
\begin{equation}
P_g/E_T  \sim 1.7 (T/10^4 K)V_{10}^{-2}.
\end{equation}
If $P_{ext}$ in a galaxy is a factor $F$ times the gas pressure, 
then the peak field given by Eq.\ (\ref{seclim})
begins to exceed that given by  Eq.\ (\ref{finlim}), when the
ion density exceeds a critical value $n_i^c$. This critical
value is given by
\begin{eqnarray}
n_i^c &\sim& 6.4 \times 10^{-2} ({n_n \over 1 cm^{-3}})^{2/3} 
{\rm cm}^{-3} \nonumber\\
&\times& V_{10}^{-3} L_{100}^{-1/3}({L\over l})^{4/3}
({F\over 4})^{7/3} ({T \over 10^4 K})^{7/3} 
 .
\label{ionlim}
\end{eqnarray}

For a larger ion density than $n_i^c$,
the peak field for the
flux ropes tangled on the largest scale $L$ saturates
to a value $B_p \sim (8\pi P_{ext})^{1/2}$, lower
than given by  Eq.\ (\ref{seclim}) . So, for the
small-scale dynamo-generated field to saturate to sub-equipartition 
level by the processes described above, 
the galactic gas has to be predominantly neutral
with an ion density less than $n_i^c$.
Note that the value of $n_i^c$ is critically dependent on the 
turbulence parameters that obtain in the ISM of the galaxy, especially
the turbulent velocity scale. For example for
$V= 5km s^{-1}$, and all other parameters as
above, the critical density becomes $n_i^c \sim 0.5 cm^{-3}$;
so even for an ionised hydrogen density as large as $10\% - 50\%$ of
a neutral density, taken here to be $ 1 cm^{-3}$), saturation
could occur due to inefficient stretching.
 
Let us now ask what happens if the ion densities
exceed the critical value $n_i^c$. In this case 
$(B_p^2/8\pi) < P_{ext}$ and 
the stretching constraint 
can only be satisfied only if the flux rope thickens further
than the value implied by ambipolar drift, to
a radius $ r_d = R \sim  l (C_d/4\pi) (\rho_n v_l^2/(2P_{ext}))$.
If the flux ropes can thicken to this radius,
the average energy magnetic energy density will be
\begin{equation}
{E_B(L) \over E_T} \sim  N {3C_d^2 \over 128\pi F} M_T^2 \sim 
7.3 \times 10^{-4} NC_d^2
\label{aneb}
\end{equation}
where $M_T$ is the Mach number of the turbulence and we
have adopted the parameters $F\sim 4$ and $P_g/E_T \sim 1.7$, 
given above for the numerical estimate. The value of
N should again be limited by the collapse of loops
as discussed above. 

However, it is not clear if
flux ropes can thicken further than the radius implied by 
ambipolar drift, when dynamo action
begins from weak seed fields. One possibility is
that the magnetic pressure in the ropes 
acting on the fluid as a whole can thicken the rope.
But this can only happen if the pressure in the 
rope becomes larger than the pressure outside, at least 
temporarily. Even if this were possible, as the flux rope thickens, 
flux freezing leads to a decrease of the field 
strength in the rope, and a consequent decrease in the thickening
rate. The fluid pressure in the rope will also decrease as the rope
expands. Note that this problem does not arise when ambipolar 
drift is causing the thickening. In a predominantly neutral
medium, the ion pressure is much smaller than that due to neutrals.
And the Lorentz force term in the Euler equation for the ions can
be much larger than the ion pressure gradient term,
and cause a relative drift
of the ions with respect to the neutrals, carrying the field, and
hence thickening the flux rope. 

If the ion density is larger than
$n_i^c$ and the flux ropes cannot thicken sufficiently to 
resist stretching, then it is not clear how exactly the small
scale field saturates. The field may
get packed locally into a radius $\sim R$ by the
folding motions associated with the turbulence, and 
then resist further stretching. On the other hand,
the small-scale field may only saturate if the length of 
flux ropes, $N$, increases sufficiently to achieve equipartition with
the turbulence. In either case the small-scale field will
be highly intermittent. The effects of reconnection
of this highly intermittent field (cf. Vishniac 1995, 
Lazarian and Vishniac 1996) could be important
in deciding if the turbulence can still lead to large-scale
dynamo action. In an interesting paper
which came to our notice during the completion
of the present work, Blackman (1996) 
discusses the possible effects of reconnection in greater detail,
albiet in the case where the flux ropes are assumed to have
a thickness $\sim R$, and assuming the gas is largely ionised.

\section{ Discussion and conclusions}

The large-scale galactic field is thought to be generated by
a turbulent dynamo. However the same turbulence will
produce magnetic noise at a more rapid rate. 
We have examined whether the 
Lorentz forces associated with the growing small-scale fields,
can lead to their saturation, in a
manner which preserves large-scale dynamo action. 
In doing this, we have also taken account of the ambipolar drift
induced by the presence of a neutral component of
the galactic gas.

The saturated state of the small-scale dynamo
generated field, which we have motivated, 
does indeed preserve large-scale dynamo action.
The crucial property of the small-scale dynamo
generated field which allows this to happen is its spatial
intermittency. The field can build up locally
to a level which will lead to small-scale dynamo saturation,
while at the same time having a sub-equipartition 
{\it average} energy density. 

Numerical simulations
of dynamo action due to mirror-symmetric turbulence
(Meneguzzi {\it et al.} 1981) or convection 
(Brandenburg {\it et al.} 1996) have indeed hinted at
a saturated state of the small-scale dynamo 
as described above; a magnetic field concentrated
into flux ropes, occupying a small fraction of the fluid volume,
having peak fields comparable or in excess of equipartition
value but average magnetic energy density only
about $10\%$ of the kinetic energy density.

We have described in section 4 and 5 the  
approach to this saturated state.  As we noted in section 4, 
for conditions appropriate to galactic gas,
the effective magnetic Reynolds number, even including ambipolar
diffusion, is much larger than a critical value needed
for small-scale dynamo action. However, in such a case,
as the the small-scale field grows in strength, it 
continues to be concentrated into thin ropy structures, 
as in the kinematic regime. 
These flux ropes are curved on the turbulent eddy scales, while their
thickness is set by the diffusive scale determined by the
effective ambipolar diffusion. 
The growing magnetic tension associated with the curved flux ropes, 
acts to straighten them out. 
Frictional drag damps the magnetic energy
associated with the wrinkle in the rope. Also, small-scale flux loops 
can collapse and disappear. These non-local effects operate on
the eddy turnover time scale, when the peak field in a flux rope 
has grown to a few times the equipartion value. Their net effect is to make
the random stretching needed for the small-scale dynamo
inefficient and hence saturate the small-scale dynamo.
However, the average energy density in the saturated small-scale field is
sub equipartition, since it does not fill the volume.
Such fields neither drain significant 
energy from the turbulence, nor convert eddy motion of the turbulence 
on the outer scale to wave-like motion. The 
diffusive effects needed for the large-scale dynamo 
operation are then preserved.
This picture of small-scale dynamo saturation obtains
when the ion density is less than a 
critical value of $n_i^c \sim 0.06 - 0.5 cm^{-3} (n_n/ cm^{-3})^{2/3}$.

For very large ion densities $n_i > n_i^c$,
the small-scale field is expected to saturate only when its
energy density grows comparable to that of the turbulence. 
This is because, in this case the peak field, cannot grow sufficiently, 
(without its pressure exceeding the interstellar pressure), 
for the stretching constraint to apply. However,
the spatial structure of the small-scale field, 
is still likely to be highly intermittent. The
large-scale dynamo action will depend on how such a field
responds to turbulent motions, especially whether the field can
reconnect efficiently (cf. Vishniac 1995, Blackman 1996).

We discuss briefly, the implications of 
the above results for the origin of galactic fields.
The viability of the saturation mechanism discussed here for limiting
magnetic noise depends on the ionisation of the gas
and the turbulence parameters. 
In the context of the Galaxy, a study of warm clouds by
Spitzer and Fitzpatrick (1993) gives a range
of electron densities for the clouds, with an average of
$\sim 0.07 cm^{-3}$. They also deduce an average neutral
density $\sim 0.2 cm^{-3}$. For such parameters, 
the magnetic noise will indeed saturate if the turbulent
velocity is $\sim 5 km s^{-1}$ and other parameters
are as in Eq. (\ref{ionlim}) . 

In the case of a young galaxy
with mass $\sim 10^{11} M_\odot$ that has
just collapsed into a $10kpc$ sized region, the average density is
larger, and is $\sim 1 cm^{-3}$. The average column density for $HI$ is
$\sim 10^{22} cm^{-2}$. The damped Ly-$\alpha$ systems
seen in the spectra of high redshift quasars, and thought
to be young galaxies do indeed have such HI column densities
(cf. Wolfe 1995). The ionisation fraction is more uncertain.
For such a high HI column density as inferred above, 
the gas is expected to become
self shielded to external ionising flux. However, ionisation will
result from UV emission from young stars embedded in the gas, 
whose importance depends
on the uncertain star formation rates and stellar mass functions.
From an observational point of view,
a recent study of metal lines in these systems (Lu {\it et al.}, 1996)
deduces an upper limit to their electron density, consistent
with the average electron density in warm clouds in the
Milky Way $\sim 0.07 cm^{-3}$ mentioned above. Further
velocity widths deduced from 21cm absorption due to neutral hydrogen
in several damped Ly-$\alpha$ systems, also limit
a turbulent component to the line width to be 
about $10 km s^{-1}$ (cf. De Bruyn {\it et al.} , 1996).
So, here also, the densities
and turbulence parameters are expected to be in the range wherein 
the small-scale dynamo generated fields 
can saturate due to the tension forces, in a 
way which preserves large-scale dynamo action.

In our analysis, so far, we have ignored the small-scale
field generated by the tangling of the large-scale
field by the turbulence. Even when the dynamo generated small
scale field has saturated, this will provide an additional
source of small-scale magnetic noise. As the large-scale 
field grows, so does this
component of the small-scale field with an energy density
ultimately decided by the nature of
the MHD turbulence (cf. Zeldovich {\it et al.} 1983).
We hope to return to this issue in a later work.
One then expects two components to the small-scale
magnetic field in the interstellar medium of a galaxy.
First, a ropy, intermittent component, with flux ropes curved on scale
$L \sim 100 pc$, say, and thickness $r_d \sim 10^{-2} L
\sim 1 pc$, with peak field a few times equipartition. Second, 
a more diffuse small-scale field related in
strength to the large-scale field. It would be interesting to 
search for both of these components in the interstellar medium
of galaxies. 

The effect of ambipolar drift, together with the small-scale 
dynamo, can also influence galactic magnetic
field generation, indirectly, in another fashion. Note that any dynamo
needs a seed field to act upon. Since the small-scale dynamo
acts to generate fields more rapidly than the large-scale
dynamo, the magnetic noise so generated may itself
provide a significant seed for the large-scale dynamo
(cf. Beck {\it et al.} 1994).
However the small-scale dynamo also leads to
highly ropy fields with a rope
thickness, $r_d$. In a fully ionised gas $r_d = r_d^i \sim L/R_m^{1/2}$
will be very small, since $R_m >> 1$. The overlap 
of such a field with a large-scale dynamo eigenfunction will
be small. However, if small-scale dynamo
action proceeds in the presence of neutrals, $r_d =
r_d^n \sim L/R_{ambi}^{1/2} >> r_d^i$ in general, since $R_{ambi} << R_m$.
So, when neutrals are present,
the small-scale dynamo-generated
magnetic noise will provide a more coherent seed field,
for large-scale dynamo action.
This will act to shorten the timescale for 
the generation of large-scale
galactic magnetic field, to the microgauss level, at higher redshift.

Note that, after recombination, the residual ionisation
fraction of the intergalactic medium drops to about
$10^{-4} - 10^{-5}$. One may  be tempted to apply 
some of the results obtained here
to the first generation of objects 
which collapse at high redshifts. 
The limitations of our semi-quantitative arguments, and 
our assumption of a homogeneous galactic interstellar medium, 
in reaching the above conclusions, needs little emphasising. 
It would also be fruitful to find ways of incorporating,
more fully, the dynamics of the velocity correlations, as
we have done for the magnetic correlations. This
full MHD turbulence problem appears formidable at present.
Nevertheless, the results obtained here encourage the belief that
that the turbulent galactic dynamo could indeed be made 
to produce large-scale fields, in the presence of a significant
neutral component.

\section*{Acknowledgments}

At Sussex KS is supported by a PPARC
Visiting Fellowship. He thanks John Barrow for invitation to Sussex
and both him and Leon Mestel for their detailed comments on the manuscript.
At Princeton KS was supported by NSF grant AST-9424416. 
Partial travel support came from IAU Commission 38.
He thanks Jerry Ostriker and Ed Turner for the invitation
to visit Princeton University Observatory, and the staff there
for a warm reception. Discussions with collegues too
numerous to mention have been very helpful, during the course of this work.

\bsp

\label{lastpage}


\begin{thebibliography}{}

\bibitem{} Beck, R., Brandenburg, A., Moss, D., Shukurov, A. $\&$
Sokoloff, D., 1996, Ann. Rev. Astron. Astrophys., 34, 155.
\bibitem{} Beck, R., Poezd, A.D., Shukurov, A. $\&$ Sokoloff, D.,
1994.. A $\&$ A, 289, 94.
\bibitem{} Brandenburg, A., 1994, {\it Lectures on Solar and
Planetary Dynamos}, ed. Proctor, M. R. E. $\&$ Gilbert, A. D.,
Cambridge University press, p117.
\bibitem{} Brandenburg, A., Jennings, R. L., Nordlund, A., 
Rieutord, M., stein, R. F. $\&$ Touminen, I., 1996. JFM, 306, 325.
\bibitem{} Blackman, E., 1996. Phys. Rev. Lett., 77, 2694.
\bibitem{} Chandran, B., 1996. Princeton University Observatory
preprint POPe-659.   .
\bibitem{} Cattaneo, F. $\&$ Vainshtein, S. I., 1991. ApJ, 376, L21.
\bibitem{} De Luca, E. E., Fisher, G. H. $\&$ Patten, B. M., 1993.
ApJ, 411, 383.
\bibitem{} De Bruyn, A. G., O'Dea, C. P. $\&$ Baum, S. A., 1996.
A $\&$ A, 305, 450.
\bibitem{} Draine, B. T., 1980. ApJ, 241, 1021.
\bibitem{} Draine, B. T., 1986. MNRAS, 220, 133.
\bibitem{} Field, G. B., 1996. {\it The Physics of the Interstellar
Medium and Intergalactic Medium}, ed. Ferrara {et al.}, ASP
Conf. Ser., 80, p1.
\bibitem{} Kazantsev, A. P., 1968. Sov. Phys. JETP, 26, 1031.
\bibitem{} Kleeorin, N. I., Ruzmaikin, A. A., Sokoloff, D. D., 1986.
{\it Plasma Astrophysics}, ESA Publication, p557.
\bibitem{} Krause, F. $\&$ Radler, K.-H., 1980. {\it Mean-Filed
Magnetohydrodynamics and Dynamo Theory}, Pergamon press, Oxford.
\bibitem{kul92} Kulsrud, R.M. $\&$ Anderson, S.W., 1992. ApJ., 396, 606.
\bibitem{} Landau, L. $\&$ Lifshitz, E. M., 1987. {\it Fluid Mechanics},
Pergamon Press, Oxford.
\bibitem{} Lazarian, A. $\&$ Vishniac, E., 1996. {\it Polarimetry of the 
Interstellar medium}, ASP Conf. Ser., 97, p537.
\bibitem{} Lu, L., Sargent, W. L. W., Barlow, T. A., Churchill, C. W.
$\&$ Vogt, S. S. 1996, ApJ Suppl., 107, 475. 
\bibitem{} Meneguzzi, M., Frisch, U. $\&$ Pouquet, A., 1981. Phys. Rev. Lett.,
47, 1060.
\bibitem{} Mestel, L. $\&$ Spitzer, L., 1956. MNRAS, 116, 503.
\bibitem{} Mestel, L. $\&$ Subramanian, K., 1991. MNRAS, 248, 677.
\bibitem{} Moffat, H. K., 1978. {\it Magnetic Field Generation
in Electrically Conducting Fluids}, Cambridge University Press, Cambridge.
\bibitem{} Parker, E. N., 1979. {\it Cosmic Magnetic Fields}, 
Clarendon press, Oxford.
\bibitem{} Ratra, B., 1992. ApJ Lett., 391, L1.
\bibitem{} Rees, M. J., 1987. Quart. J.R.A.S., 28, 197.
\bibitem{} Rees, M.J., 1994. {\it Cosmical Magnetism}, 
ed. Lynden-Bell, D., Kluwer, London, p155 
\bibitem{} Ruzmaikin, A. A., Shukurov, A. M. $\&$ Sokoloff, D. D., 1988.
{\it Magnetic Fields of Galaxies}, Kluwer, Dordrecht.
\bibitem{} Ruzmaikin, A. A., Shukurov, A. M. $\&$ Sokoloff, D. D., 1989.
MNRAS, 241, 1.
\bibitem{} Spitzer, L., 1978. {\it Physical Processes in the
Interstellar medium}, Wiley-Interscience, New York.
\bibitem{} Spitzer, L. $\&$ Fitzpatrick, E., 1993. ApJ, 409, 299.
\bibitem{} Subramanian, K., 1995. Bull. Astr. Soc. Ind., 23, 481.
\bibitem{} Subramanian, K., 1997. in Preparation (Paper II).
\bibitem{} Subramanian, K., Narasimha, D. $\&$ Chitre, S.M., 
1994. MNRAS. 271, L15.
\bibitem{} Vainshtein, S. $\&$ Kichatinov, L. L., 
1986, J. Fluid. Mech., 168, 73.
\bibitem{} Vainshtein, S. $\&$ Rosner, R., 1991, ApJ., 376, 199.
\bibitem{} Vishniac, E. T., 1995. ApJ, 446, 724.
\bibitem{} Wolfe, A. M., 1995. {\it QSO Absorption Lines}, ed. 
Meylan G., Springer, Berlin, p13.
\bibitem{} Zeldovich, Ya.B., Ruzmaikin,A.A. $\&$ Sokoloff,D.D., 1983. {\it
Magnetic fields in Astrophysics}, Gordon and Breach, New York.
\bibitem{} Zweibel, E. G., 1986. ApJ, 329, 384.
 

\end{thebibliography}
\end{document}